\documentclass[3p,times,procedia]{elsarticle}
\usepackage{nupha_ecrc}
\usepackage[utf8]{inputenc}

\volume{00}

\firstpage{1}

\journalname{Nuclear Physics A}

\runauth{}

\jid{nupha}

\jnltitlelogo{Nuclear Physics A}

\usepackage{amssymb}

\usepackage{lineno}

\usepackage[figuresright]{rotating}

\begin{document}

\begin{frontmatter}



\dochead{XXVIIIth International Conference on Ultrarelativistic Nucleus-Nucleus Collisions\\ (Quark Matter 2019)}

\title{Probing QGP with flow: An experimental overview}


\author{Katar\'ina K\v r\'i\v zkov\'a Gajdo\v sov\'a}

\address{Czech Technical University in Prague, Faculty of Nuclear Sciences and Physical Engineering, B\v rehov\'a 7, 115 19, Prague, Czech Republic}

\begin{abstract}
An experimental overview of anisotropic flow measurements and their ability to probe the properties and the nature of the system created in ultra-relativistic hadron collisions is given in these proceedings. The aim is to discuss the state-of-the-art measurements ranging from small to large systems at different collision energies.
\end{abstract}


\end{frontmatter}


\section{Introduction}
\label{sec:introduction}

Relativistic Heavy Ion Collider (RHIC) at BNL and the Large Hadron Collider (LHC) at CERN are machines standing at the forefront of the research of a hot and dense, strongly interacting QCD medium, called the quark-gluon plasma (QGP). One of their main purposes is to recreate the QGP via ultra-relativistic heavy-ion collisions in order to study its properties at extreme conditions, using collisions of smaller systems (such as p-A or pp) as a reference with absence of such medium. It is known for a couple of years, that measurements in small collision systems with high particle multiplicities reveal features similar to those observed in the collectively expanding medium present in heavy-ion collisions.

In collisions of heavy ions, spatial anisotropy of their overlap region in the transverse plane is translated to an anisotropic distribution of final-state particles via parton interactions during the deconfined phase. The azimuthal distribution of emitted particles can be decomposed into Fourier series as
\begin{equation}
P(\varphi) = \frac{1}{2\pi} \sum_{n=-\infty}^{\infty}V_n e^{-in\varphi},
\end{equation}
where $v_n$ is the magnitude (also called the flow coefficient) and $\Psi_n$ is the phase (also called the symmetry plane) of the flow vector $V_n = v_ne^{in\Psi_n}$. The values of flow vectors reflect the hydrodynamic response of the medium to the initial state eccentricity.
Thus, measurements of flow vectors, their fluctuations and/or correlations provide an important ingredient for validation of the existing theoretical models, in particular for determination of initial conditions and transport coefficients of the medium~\cite{Gale:2013da,Heinz:2013th,Song:2017wtw}.

In small collision systems, the same observables are used to resolve, whether a medium of a similar origin as in heavy-ion collisions is formed. As opposed to heavy-ion collisions, it is not yet clear, to what extent are the flow vectors driven by the initial spatial anisotropy, and whether the gluon field momentum correlations from the initial state persist and contribute to the observed final state anisotropy~\cite{Dusling:2015gta,Nagle:2018nvi}. 
In addition, an overwhelming contamination from non-flow effects, arising mainly from correlations of particles within jets, leads to inevitable complications in measurements performed in small systems. A template fit method used in measurements of two-particle correlation functions~\cite{Aad:2015gqa} and the subevent method used in two- and multi-particle cumulants~\cite{Jia:2017hbm} are the state-of-the-art approaches to suppress contributions from non-flow. Any interpretation of measurements in small systems should not be advanced without appropriate treatment of this contamination.

Since the debates about large and small collision systems differ in their essence, the measurements presented here will be  separated into two sections. Yet, the discussion about small systems is often accompanied by references and comparisons with the collective AA collisions. I would like to note that this is not an exhaustive summary of the results presented at the conference, but rather a general overview of the latest experimental developments in the field. For a theory overview please refer to Ref.~\cite{Shen:2020gef}.

\section{Large collision systems}
\label{sec:LargeSystems}

Our understanding of the QGP has improved significantly with the measurements of magnitudes of flow vectors, $v_n$~\cite{Gale:2013da,Heinz:2013th,Song:2017wtw}. 
However, details of the initial conditions and/or the dynamics of the subsequent deconfined phase cannot be resolved with measurements of $v_n$ alone. Fortunately, a wealth of experimental data collected over the past years allow us to dive deeper into the investigations of the QGP and improve our knowledge about this phase of QCD matter.

Event-by-event fluctuations of the initial state geometry cause the flow vectors (their magnitudes and symmetry plane angles), constructed in different $p_T$ or $\eta$ ranges, to fluctuate around the event-averaged values. Investigations of these fluctuations pose important constraints on initial conditions, which in turn contribute to more precise modeling of the final state dynamics.
This can be addressed with several measurements, such as the event-by-event fluctuations of the $v_n$, decorrelations of flow vectors in $p_T$ and $\eta$, and in performing the event shape engineering (ESE). A selection of the most recent developments in each of these measurements is discussed below.

\begin{figure}[!htb]
\centering
\begin{minipage}{0.5\textwidth}
\centering
\includegraphics[width=0.75\textwidth]{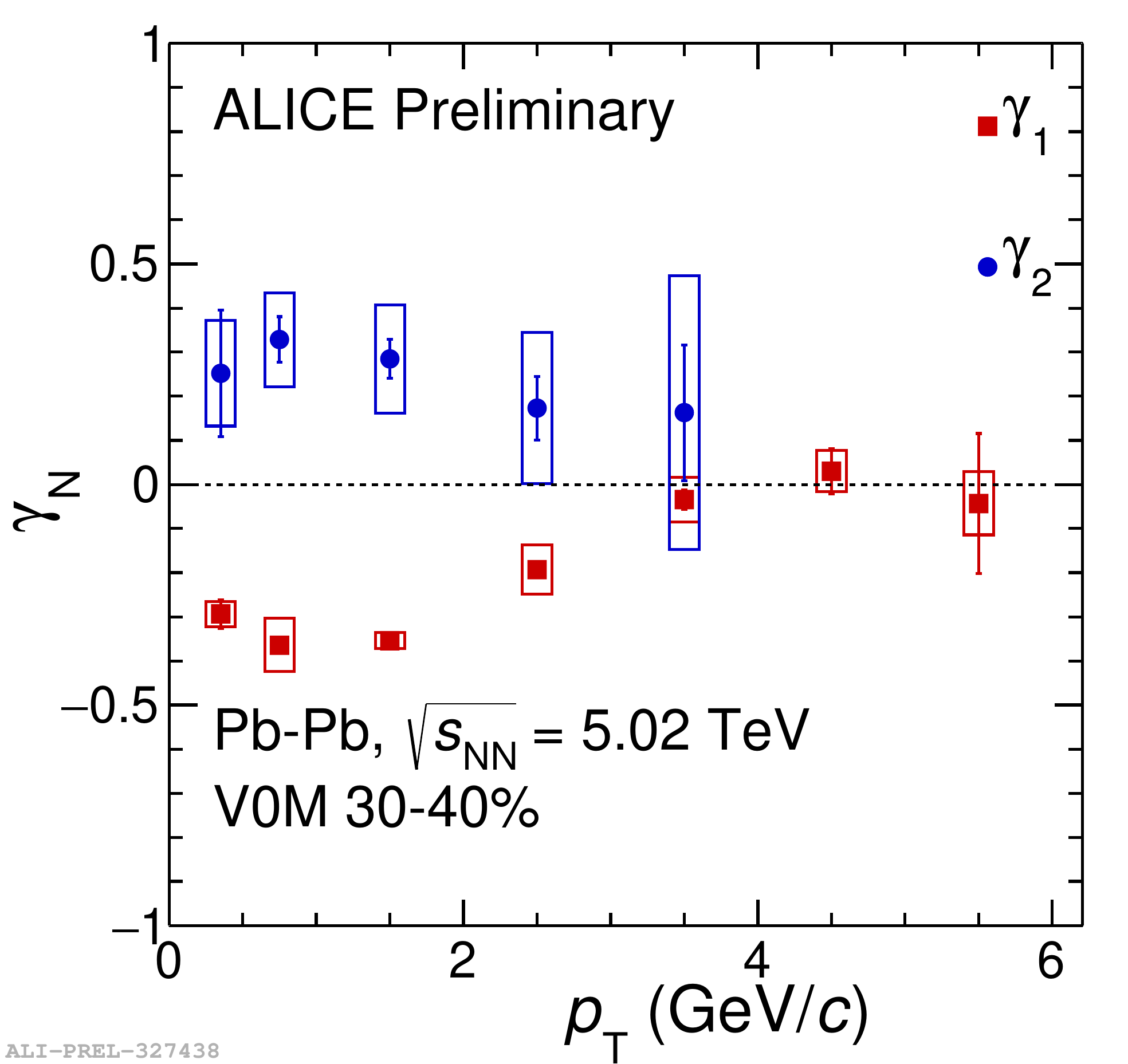}
\end{minipage}\hfill
\begin{minipage}{0.5\textwidth}
\centering
\includegraphics[width=0.95\textwidth]{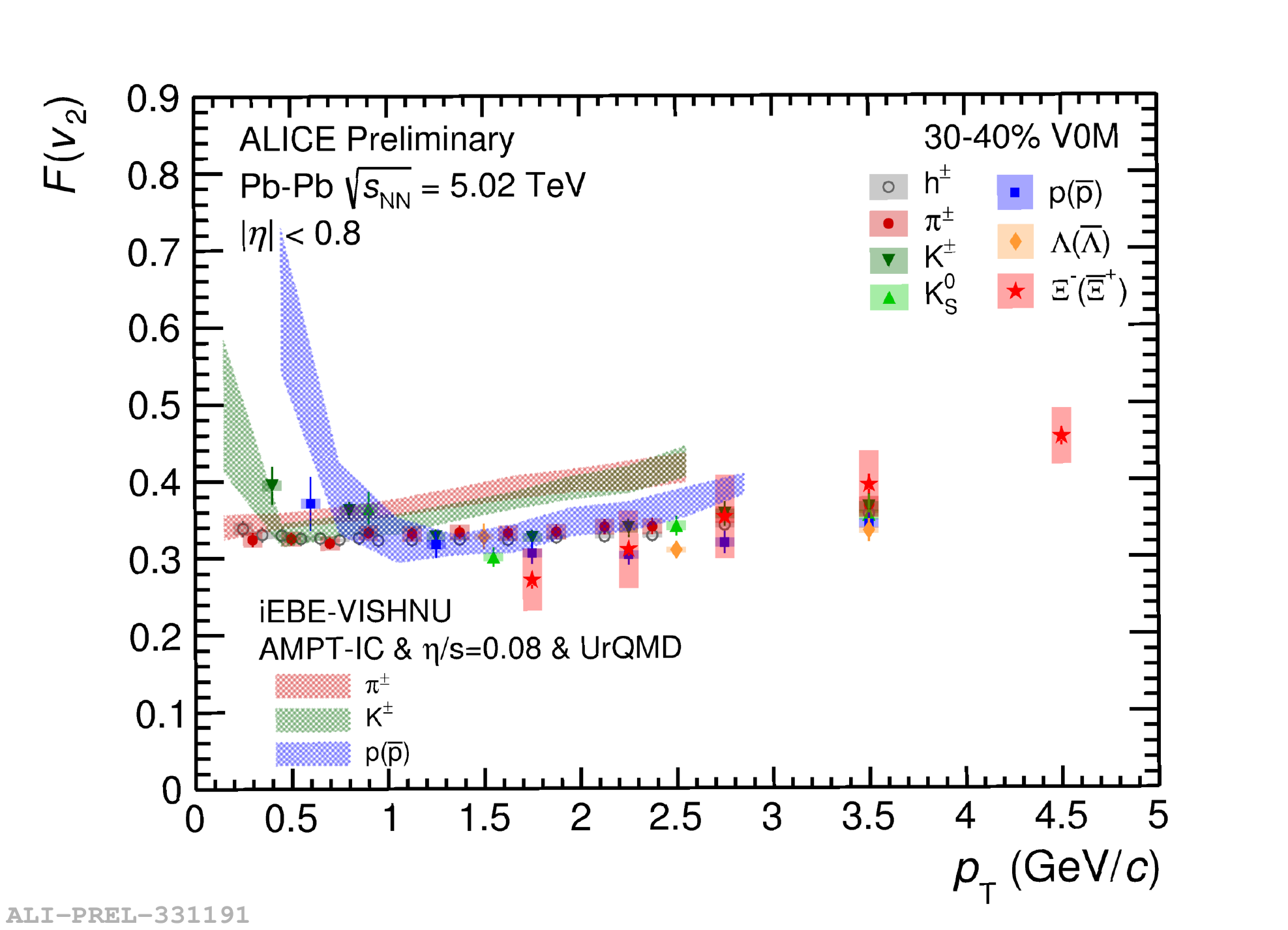}
\end{minipage}
\caption{Left: Skewness $\gamma_1$ and kurtosis $\gamma_2$ of the flow p.d.f. measured as a function of $p_{\rm T}$ in Pb--Pb collisions at $\sqrt{s_{\rm NN}} = 5.02$ TeV~\cite{yazhu}. Right: Relative flow fluctuations of identified hadrons measured as a function of $p_{\rm T}$ in Pb--Pb collisions at $\sqrt{s_{\rm NN}} = 5.02$ TeV~\cite{yazhu}, compared to iEBE-VISHNU model with AMPT initial conditions~\cite{Zhao:2017yhj}.}
\label{fig1}
\end{figure}

The flow probability density function (p.d.f.), the $P(v_n)$, can be accessed either via the unfolding procedure as used in~\cite{Aad:2013xma,Sirunyan:2017fts}, or by investigating the degeneracy of higher order cumulants by measuring deviations of their ratios from unity~\cite{Voloshin:2007pc,Giacalone:2016eyu}. Measurements of inclusive charged hadrons revealed that flow fluctuations are neither Gaussian, nor Bessel-Gaussian~\cite{Aad:2013xma, Sirunyan:2017fts,Acharya:2018lmh, Aaboud:2019sma}. Impressive advancement in the collected data at the LHC and in the analysis techniques allowed to study flow fluctuations differentially in $p_T$. The observed $p_T$-dependence of multi-particle cumulant ratios, skewness and kurtosis, shown in Fig.~\ref{fig1} (left), indicates influence of final state fluctuations~\cite{yazhu}. 
While deviations from a Bessel-Gaussian flow p.d.f. were confirmed at $p_T<3$ GeV/$c$, degeneration of multi-particle cumulants found at intermediate $p_{\rm T}$ indicates a recovery of this parametrisation. First measurements of the four-particle cumulant $v_2\{4\}(p_T)$ of identified particles allowed to study their relative flow fluctuations via the $F(v_n) = \sigma_{v_n}/\langle v_n\rangle$ (see Fig.~\ref{fig1} (right))~\cite{yazhu}. While the iEBE-VISHNU hydrodynamic model~\cite{Zhao:2017yhj} able to reproduce the $v_n$ measurements, predicts a particle species dependence of flow fluctuations, data seem to disfavor this result, showing their potential for further constraints on theoretical calculations. Finally, it should be noted that flow fluctuations were found to be affected by the so-called volume fluctuations within a fixed centrality bin, which arise from variations of the sources used to determine the event centrality~\cite{Aaboud:2019sma}. Since values of $v_n$ change with collision centrality, such effects will naturally lead to additional flow fluctuations, manifested e.g. by a ``wrong'' (positive) sign of the four-particle cumulant. The effect was assumed to be most pronounced in ultra-central collisions. However, it was found that volume fluctuations may affect the results up to mid-central collisions~\cite{Aaboud:2019sma}.

In addition to the standard centrality selection, the method of event shape engineering (ESE) imposes further selection based on the event-by-event variation of $v_n$ within a fixed centrality bin. Measurements performed with this method can help to further understand the initial conditions of a heavy-ion collision. Results of inclusive charged hadrons showed that indeed a modification of anisotropic flow is achieved by employing the ESE technique~\cite{Adam:2015eta}. New measurements of identified hadrons confirm these observations~\cite{mihaela}, and extend them with a selection on triangularity. The results show no dependence on $p_T$ and particle type. 

\begin{figure}[!htb]
\centering
\begin{minipage}{0.5\textwidth}
\centering
\includegraphics[width=0.75\textwidth]{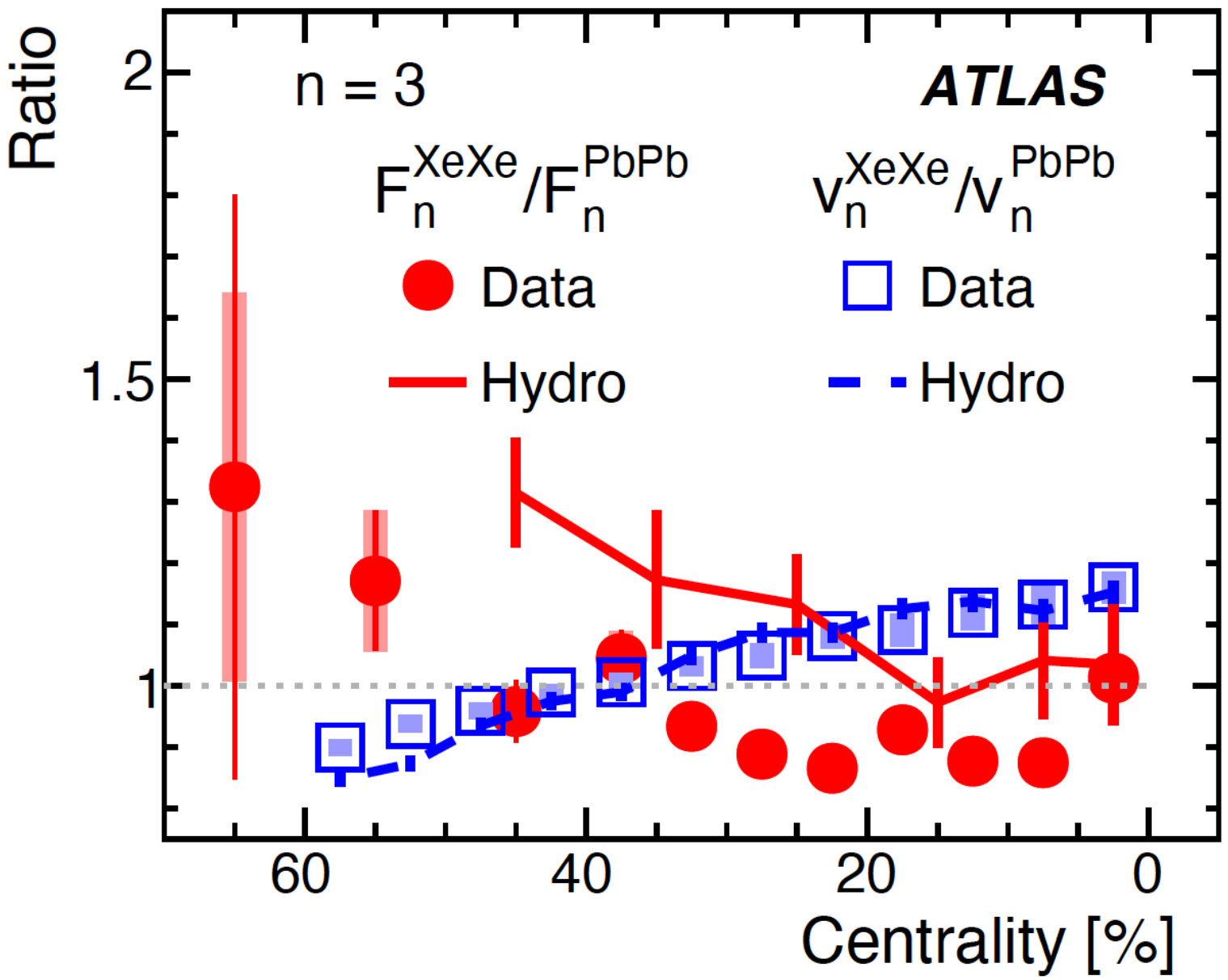}
\end{minipage}\hfill
\begin{minipage}{0.5\textwidth}
\centering
\includegraphics[width=0.7\textwidth]{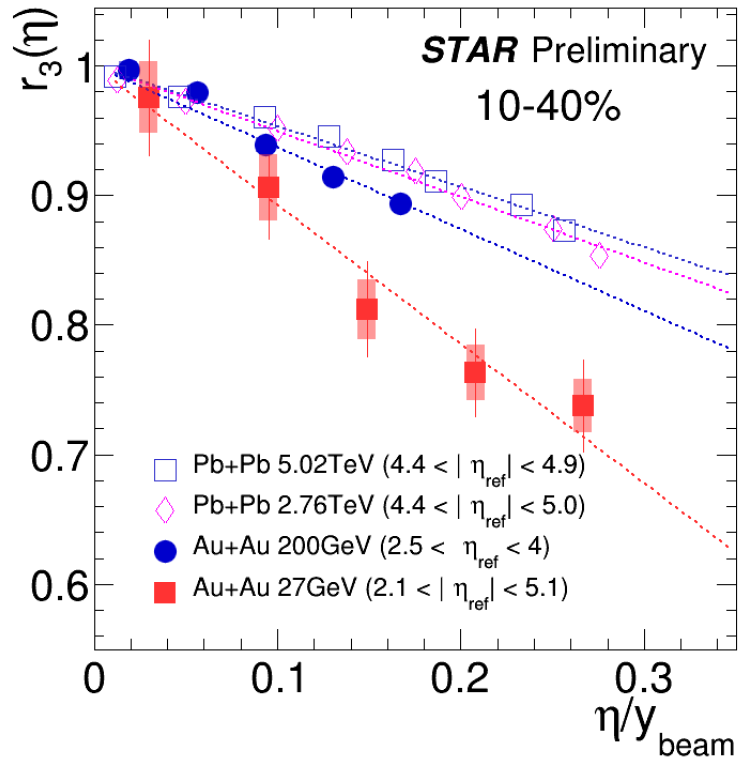}
\end{minipage}
\caption{Left: Ratios of flow decorrelations ($F_3$-ratio) and $v_3$ coefficients from Xe--Xe to Pb--Pb collisions at $\sqrt{s_{\rm NN}} = 5.44$ and 5.02 TeV, respectively~\cite{Aad:2020gfz}, compared to hydrodynamic model~\cite{Pang:2018zzo,Wu:2018cpc}. Right: Factorisation ratio $r_3(\eta)$ from Au--Au collisions at $\sqrt{s_{\rm NN}} = 200$ GeV and 27 GeV~\cite{STARdecorrelation} compared to $r_3(\eta)$ from Pb--Pb collisions at $\sqrt{s_{\rm NN}} = 2.76$ TeV and 5.02 TeV~\cite{Khachatryan:2015oea,Aaboud:2017tql}. }
\label{fig2}
\end{figure}

Flow vector decorrelations were extensively studied with the ratio $\frac{v_n\{2\}}{v_n[2]} (p_{\rm T})$~\cite{Acharya:2017ino}, or the factorisation ratio $r_n (p_{\rm T}, \eta)$~\cite{Acharya:2017ino,Khachatryan:2015oea,Aaboud:2017tql}. Absence of decorrelation would yield ratios equal to 1, while decorrelation of $v_n$ and/or $\Psi_n$ would cause the ratios to deviate from unity. These measurements can provide important constraints on the fluctuation driven dynamics of the medium, especially its longitudinal structure needed for further improvements of three-dimensional hydrodynamic models. Collisions of smaller nuclei, or collisions at lower energies, are more influenced by event-by-event fluctuations. Comparison of $r_n$ in Pb--Pb and Xe--Xe collisions recently presented in~\cite{Aad:2020gfz} provides additional sensitivity to the fluctuating initial geometry and viscous corrections. Hydrodynamic model~\cite{Pang:2018zzo,Wu:2018cpc}, tuned to describe the $v_n$ in both Xe--Xe and Pb--Pb collisions, fails to reproduce the $r_n$, as can be seen in Fig.~\ref{fig2} (left) for the $3^{rd}$ hadrmonic. Measurements of $r_n$ in Au--Au collisions at the new energy $\sqrt{s_{\rm{NN}}} = 27$ GeV~\cite{STARdecorrelation} provide further important input to theoretical modelling of heavy-ion collisions. In particular, it was found that longitudinal decorrelation of $r_3$ is stronger at lower energy (see Fig.~\ref{fig2} (right)). This effect is much more pronounced for $r_3$ in comparison to $r_2$, probably due to larger sensitivity of higher order flow harmonics to the fluctuating initial state. These, and additional results from the RHIC BES program, may significantly contribute to the advancements in the field.

\begin{figure}[!htb]
\centering
\begin{minipage}{0.5\textwidth}
\centering
\includegraphics[width=0.85\textwidth]{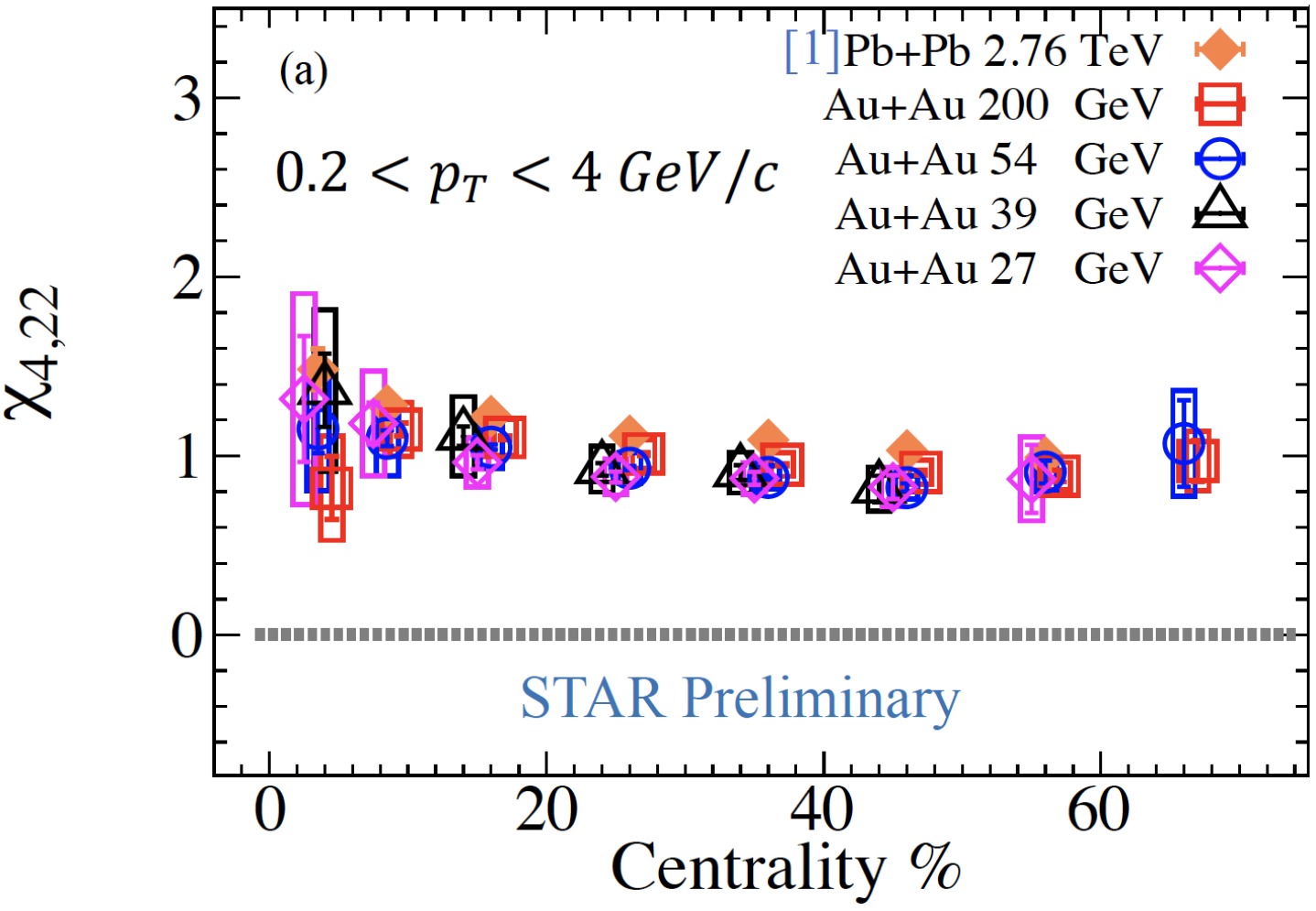}
\end{minipage}\hfill
\begin{minipage}{0.5\textwidth}
\centering
\includegraphics[width=0.8\textwidth]{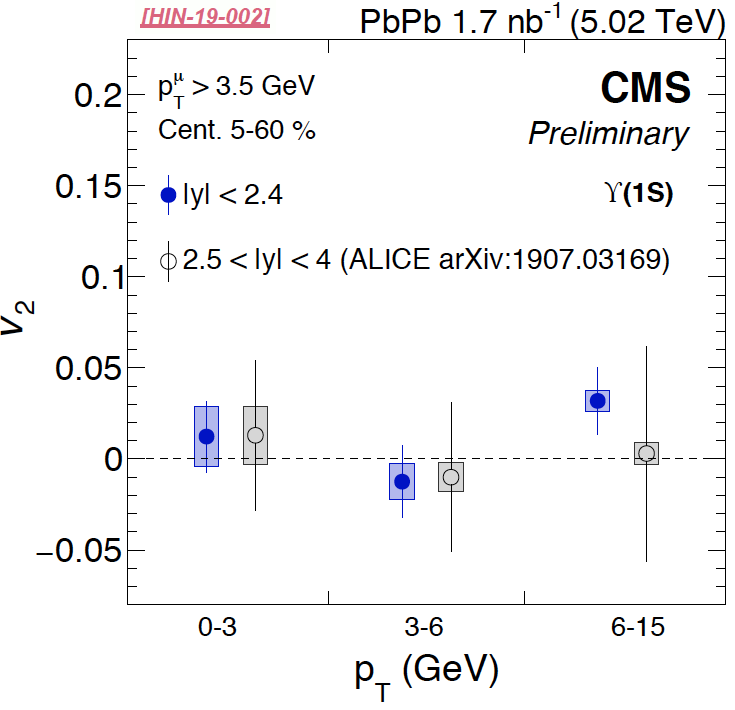}
\end{minipage}
\caption{Left: Non-linear response coefficient $\chi_{4,22} = v_4^{\rm{NL}} /\sqrt{} \langle v_2^4\rangle$ measured in Au--Au collisions at various collision energies~\cite{niseem}. Right: Transverse momentum dependence of $v_2$ of $\Upsilon$(1S) measured in Pb--Pb collisions at $\sqrt{s_{\rm NN}} = 5.02$ TeV~\cite{Acharya:2019hlv,jaebeom}.}
\label{fig3}
\end{figure}

Studies of higher order $V_n$, in particular their linear and non-linear modes, and correlations between different orders of flow coefficients or symmetry planes, provide a detailed insight into the hydrodynamic response of the system to the initial density profiles~\cite{Aad:2014fla,Aad:2015lwa,ALICE:2016kpq,Acharya:2017zfg,STAR:2018fpo}.
The main assumption in such studies lies in a linear response of $V_2$ and $V_3$ to the initial eccentricities~\cite{Alver:2010gr}, while higher order $V_n$ can be expressed in term of linear and non-linear modes, each being proportional to the same order eccentricity or lower order eccentricities and/or their products, respectively~\cite{Bhalerao:2014xra,Yan:2015jma}. 
Particularly strong constraints to models of the hydrodynamic phase can be imposed by measurements of non-linear response coefficients $\chi_{n,m,k}$. They are sensitive to the shear viscosity over entropy density ratio $\eta/s$ at freeze-out, which cannot be addressed by any other measurement performed so far~\cite{Yan:2015jma,Qian:2016fpi}.
A significant progress has been made in this direction.
Results of non-linear modes of $V_n$ extending to harmonics of high orders showed that none of the model calculations used for comparison was able to simultaneously describe all the presented observables~\cite{Sirunyan:2019izh,Acharya:2020taj}. In particular, the $\chi_{7,223}$ revealed a remarkable ability to further constrain both initial conditions, and transport coefficients of the medium~\cite{Sirunyan:2019izh}. A new testing ground for modeling of heavy-ion collisions was recently provided by measurements of non-linear flow modes of identified hadrons~\cite{Acharya:2019uia} and by the first results of energy and system size dependence of $\chi_{4,22}$ and symmetry planes correlation $\rho_{4,22}$~\cite{niseem}. As can be seen in Fig.~\ref{fig3} (left), no variation with collision energy is found for $\chi_{4,22}$.

Since the first measurements of a finite flow of heavy flavor particles~\cite{Abelev:2013lca,Adamczyk:2017xur} it has become evident that such measurements opened a new window for theory validation due to the early formation of heavy flavor particles and their subsequent participation in the collective expansion of the system.
Investigating flow of open heavy flavor particles or quarkonia at low $p_T$ can give further insight into the way of how heavy quarks interact with the medium created in heavy-ion collisions. The increasing amount of collected data both at RHIC and the LHC now allows to measure the flow of heavy flavor particles with unprecedented precision~\cite{sanghoon,tang,kelsey,todoroki}. Measurements of open heavy flavor hadrons confirm a significant non-zero $v_2$, with the possibility to disentangle contributions from bottom quark, which is found to flow with less strength than the charm quark~\cite{sanghoon,todoroki}. 
Flow of quarkonia is of particular interest to study the genuine interaction of charm and beauty with the medium. While $J/\Psi$ exhibits a significant flow signal~\cite{Acharya:2017tgv,Aaboud:2018ttm}, the $v_2$ of $\Upsilon$(1S) measured for the first time by ALICE~\cite{Acharya:2019hlv} and CMS~\cite{jaebeom} does not reveal any signs of a non-zero $v_2$ (for a comparison of the two measurements, see Fig.~\ref{fig3} (right)). Further investigation on this topic is of great interest.

\section{Small collision systems}
\label{sec:SmallSystems}

Small collision systems with large number of produced particles have attracted a lot of attention since the first appearance of the near-side ridge in the year 2010~\cite{Khachatryan:2010gv}. Whether this is a manifestation of a collective behaviour of the system created in such collisions was the main question, which required tremendous improvements in experimental techniques (especially in suppressing non-flow contamination as mentioned in section~\ref{sec:introduction}) to be answered. Multi-particle correlations spanning (very) longe range in pseudorapidity,  indicating presence of collectivity, were observed by several experiments down to the smallest collision systems and to very low energies~\cite{Khachatryan:2016txc,Aidala:2017ajz,Aaboud:2017blb,Adam:2019woz,Acharya:2019vdf,yuko}. The origin of this apparent collectivity is however still not resolved. Do we create a strongly interacting fluid-like medium, or rather a dilute system where partons undergo only few scatterings? Is flow generated as a response to the initial geometry via interactions in the final state, similarly as in heavy-ion collisions? To which extent do initial momentum correlations influence the observed flow signal? These and many more questions are still not clearly answered and require attention from both experimental and theoretical side. 

\begin{figure}[!htb]
\centering
\includegraphics[width=0.85\textwidth]{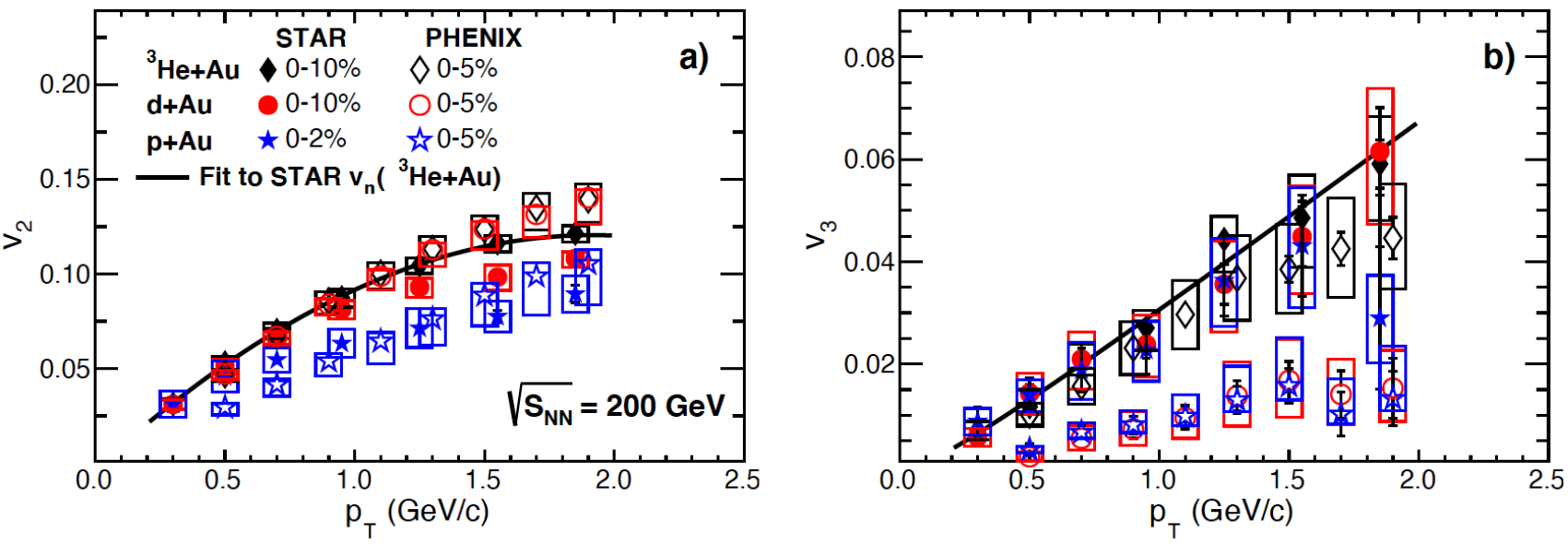}
\caption{Measurements of $v_2$ (left) and $v_3$ (right) in p--Au, d--Au and $^3$He--Au collisions at $\sqrt{s_{\rm NN}} = 200$ GeV~\cite{Lacey,PHENIX:2018lia}.}
\label{fig4}
\end{figure}

Many features of the results obtained from small systems suggest that final state effects may be responsible for our observations. One of the key findings in pA collisions is the ordering of $v_2$ depending on the mass of the studied particle~\cite{Khachatryan:2016txc,Adare:2017wlc,Pacik:2018gix}. This is consistent with observations in heavy-ion collisions, explained within a hydrodynamic picture as a result of a collective radial expansion of the medium. Final state scenario is further supported by the measurements of ratios of multi-particle cumulants which appear to be driven by the initial-state geometry~\cite{Sirunyan:2019pbr}, in particular its fluctuations instead of an overall shape. Indeed, subnucleon fluctuations were found to be crucial for a correct description of the results in small collision systems by hydrodynamic models~\cite{Mantysaari:2017cni,Weller:2017tsr}. This was recently confirmed by $v_n$ measurements in p-Au, d-Au and $^3$He-Au collisions, in particular by observing no dependence of $v_3$ on collision systems geometry~\cite{Lacey}, as opposed to the system-dependent results reported in~\cite{PHENIX:2018lia}. A comparison of the $v_2$ and $v_3$ measurements from STAR and PHENIX experiments is shown in Fig.~\ref{fig4}. While pA collisions tend to support a scenario of strong final state interactions dependent on the fluctuating initial geometry, it is not so clear in pp collisions. Until now, only few hints toward a final state description were provided by an indication of a mass ordering~\cite{Khachatryan:2016txc} and hydrodynamic description of charged particle $v_n$~\cite{Weller:2017tsr}. Nevertheless, these are challenged by several other observations, such as the inability of a full hydrodynamic simulation to translate a negative $c_2^{\epsilon}\{4\}$ to a negative $c_2\{4\}$~\cite{Zhao:2020pty}, or by alternative explanations with just few parton scatterings in a dilute system within the transport model~\cite{Kurkela:2018ygx}, or the string shoving mechanism in the PYTHIA model~\cite{Bierlich:2017vhg}.

\begin{figure}[!htb]
\centering
\begin{minipage}{0.5\textwidth}
\centering
\includegraphics[width=0.9\textwidth]{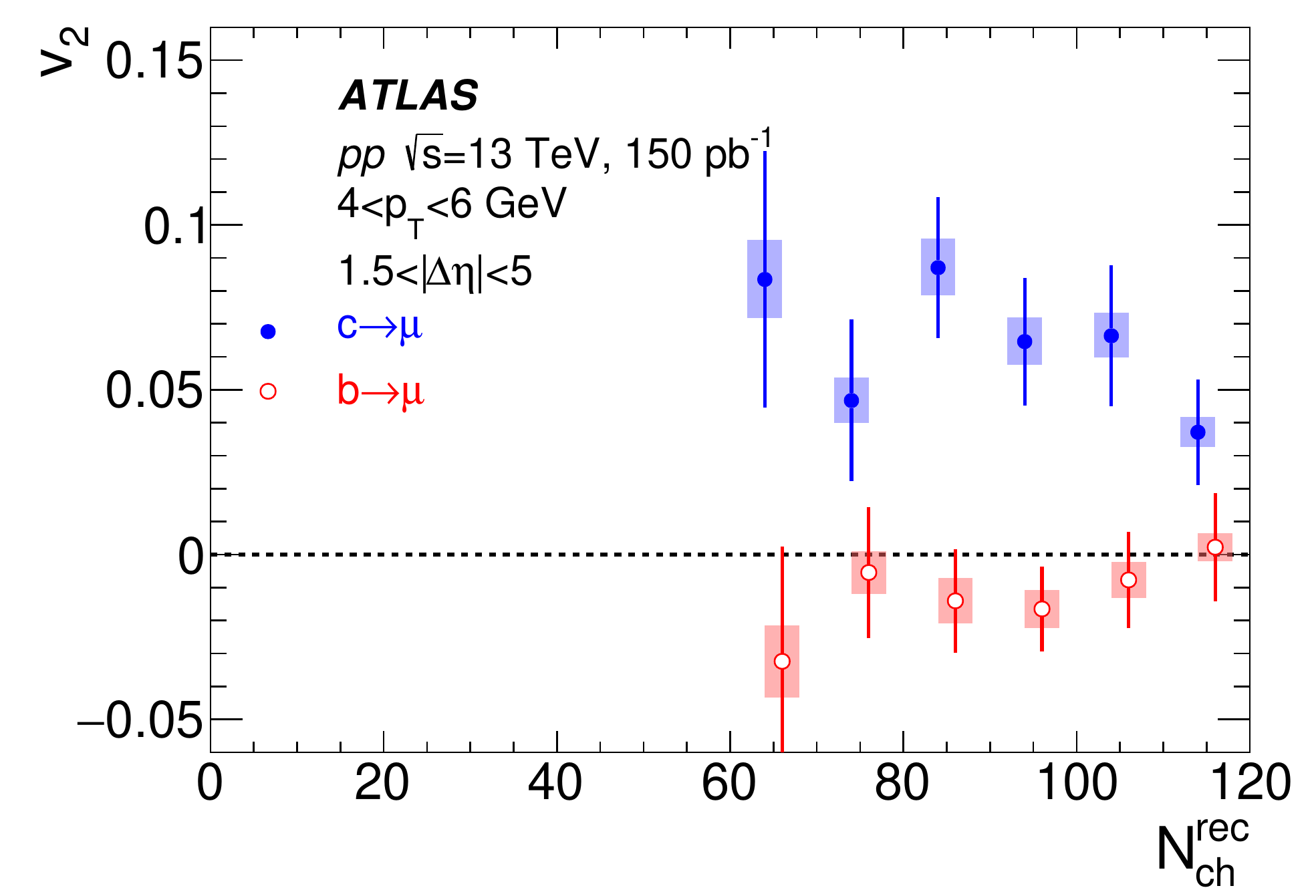}
\end{minipage}\hfill
\begin{minipage}{0.5\textwidth}
\centering
\includegraphics[width=0.85\textwidth]{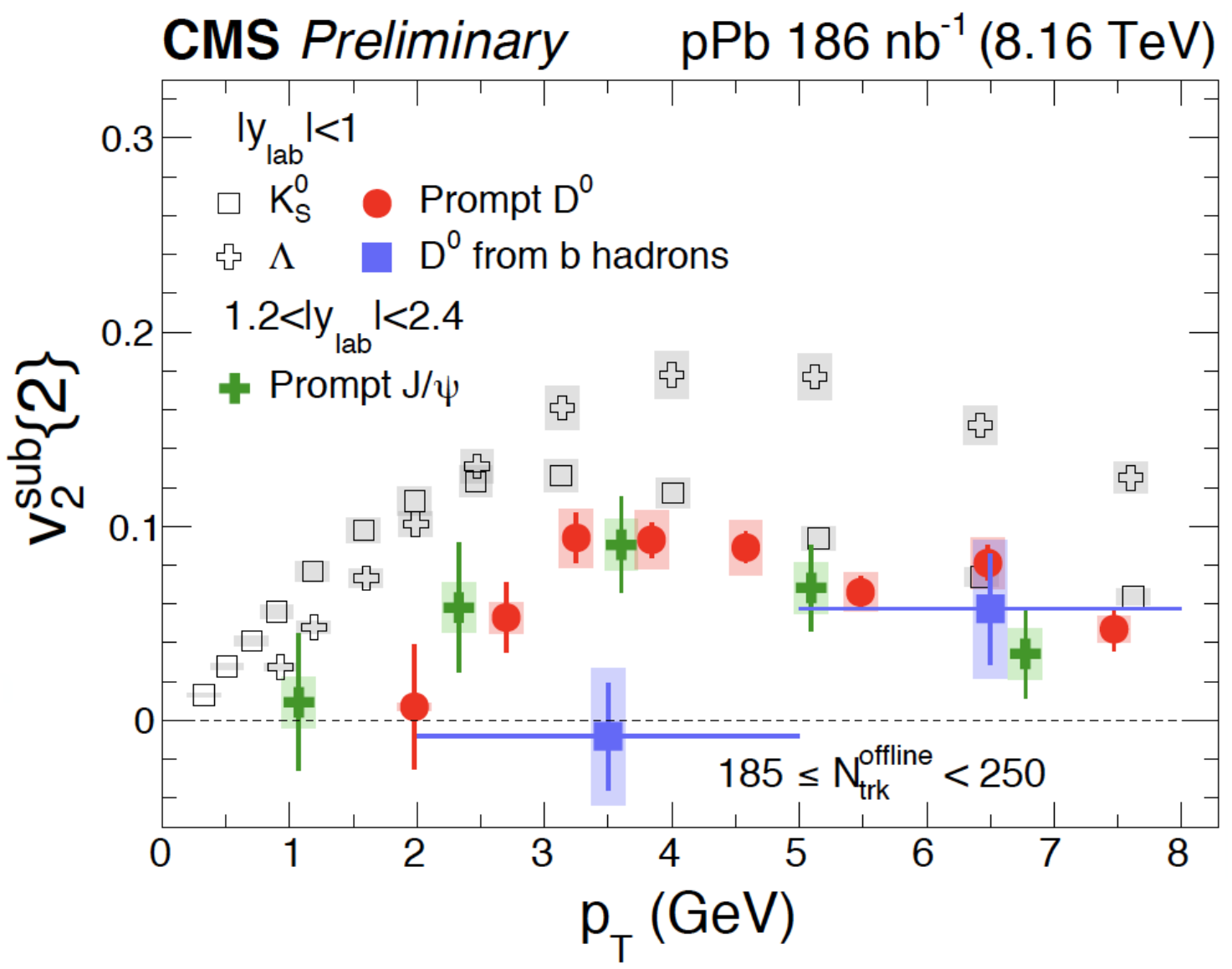}
\end{minipage}
\caption{Left: Elliptic flow $v_2$ of muons originating from charm or bottom quark as a function of $N_{\rm ch}^{\rm rec}$ measured in pp collisions at $\sqrt{s} = 13$ TeV~\cite{Aad:2019aol}. Right: Elliptic flow of non-prompt $D^0$ originating from decays of hadrons containing a $b$ quark measured in p--Pb collisions at $\sqrt{s_{\rm NN}} = 8.16$ TeV~\cite{baty}.}
\label{fig5}
\end{figure}

Even though a general consensus on the origin of collectivity has not been reached yet, the results presented here tend to support the idea of a final state scenario with a small sized fluid being created in small systems, at least in pA collisions. However, it should be kept in mind that considering the small size and short living time of the (possibly) created medium, influence of correlations from the initial state should not be neglected. It was discussed already few years ago~\cite{Greif:2017bnr} and studied again recently~\cite{Nie:2019swk}. It seems reasonable to rather focus on finding the relative balance of the two different approaches (initial vs. final state scenario), or finding a place at which one overwhelms the other. One way to study this may be the anisotropic flow of heavy flavor quarks in small systems. Since these quarks are created at very early times of a collision, they can offer a unique opportunity to disentangle the contribution from initial state correlations to the measurements of anisotropic flow. Increasing quality of the collected data at the LHC allowed to measure flow of heavy flavor particles in both p--Pb and pp collisions~\cite{tang,Aad:2019aol,baty,Acharya:2017tfn,Sirunyan:2018kiz,Acharya:2018dxy}. Recent measurements showed that while charm quark is observed to have a significant $v_2$, comparable to that of light quarks, the bottom quark did not exhibit flow~\cite{Aad:2019aol,baty}, as can be seen in Fig.~\ref{fig5}. This is in contrast to large collision systems, where similar measurements revealed a non-zero $v_2$ (it is only the bottomonium that doesn't exhibit flow, as discussed in section~\ref{sec:LargeSystems}).

\begin{figure}[!htb]
\centering
\includegraphics[width=0.45\textwidth]{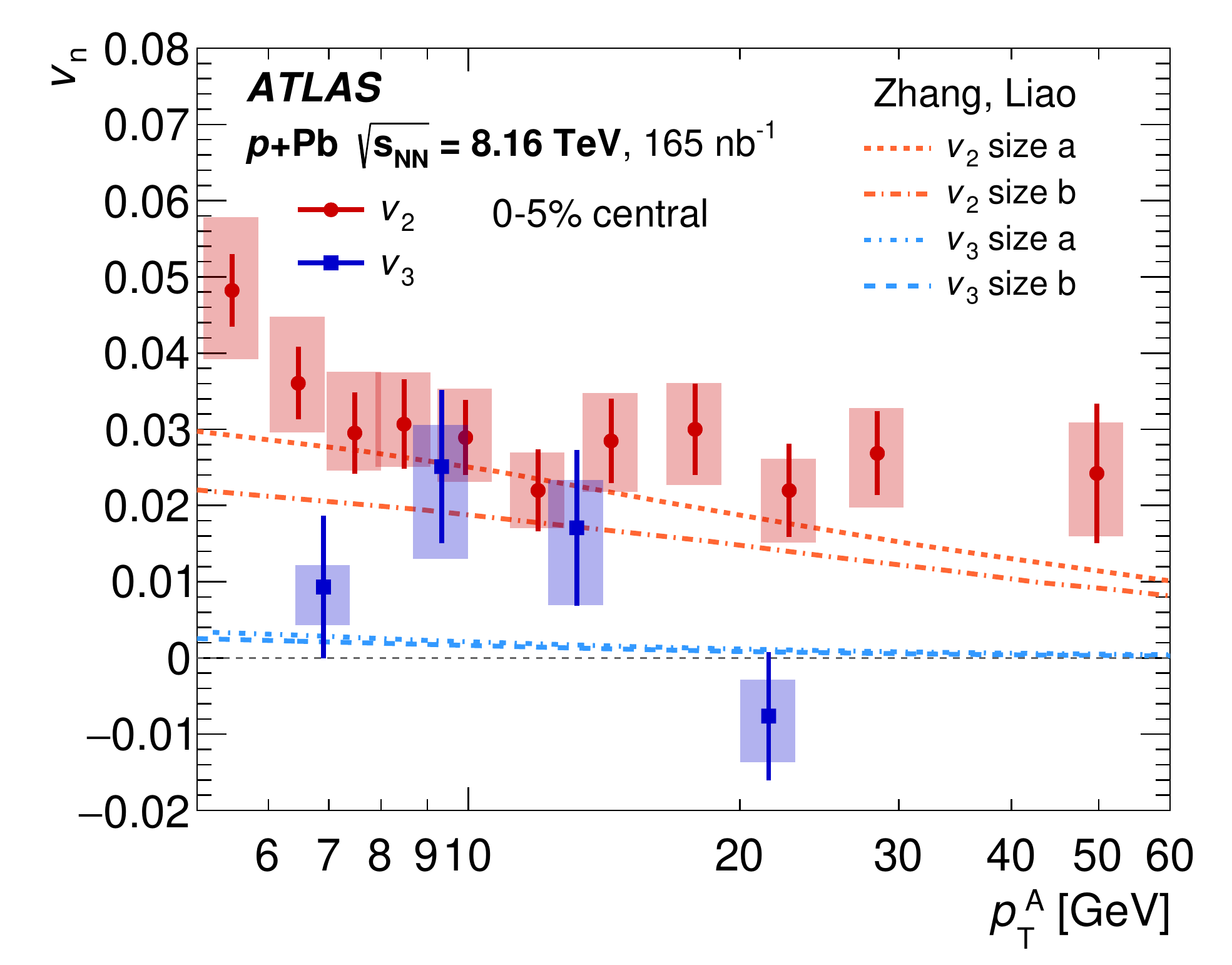}
\caption{Measurement of $v_2$ and $v_3$ at very high $p_{\rm T}$ in p--Pb collisions at $\sqrt{s_{\rm NN}} = 8.16$ TeV~\cite{Aad:2019ajj}, compared to theoretical calculations~\cite{Zhang:2013oca}.}
\label{fig6}
\end{figure}

One of the main drawbacks of the theory of having a medium similar to that created in heavy-ion collisions, is the absence of jet quenching. Until now, no apparent jet modification by the medium was found~\cite{Adam:2014qja}, as opposed to AA collisions, where the nuclear modification factor $R_{\rm AA} < 1$. A possibility of a bias in the way the normalisation of the modification ratio is obtained from simulations is discussed in~\cite{Morsch:2017brb}. Measurements of anisotropic flow at high $p_T$ offer a different way of studying the parton energy loss with the advantage of absence of such biases. As can be seen in Fig.~\ref{fig6}, finite values of $v_2$ and $v_3$ were measured at high $p_T$ of high-multiplicity p--Pb collisions~\cite{Aad:2019ajj}, which would suggest a path length dependence of the parton energy loss in a medium. Indeed, hydrodynamic model~\cite{Zhang:2013oca} invokes a  strong parton coupling to the medium in the attempt to reproduce the measurements. However, it also unavoidably leads to rather strong suppression $R_{pPb} < 1$, which is in contrast to the findings from experimental measurements~\cite{Adam:2014qja,Aad:2016zif}. These findings therefore suggest, that the observed finite azimuthal anisotropy at high $p_T$ must originate from a yet unknown mechanism. Further investigations in the direction of simultaneous description of finite $v_n$ and lack of suppression in $R_{\rm AA}$ at high $p_T$ are therefore desirable.

\section{Summary}
\label{sec:summary}

The wealth of experimental results summarised in section~\ref{sec:LargeSystems} demonstrates the level of precision that the research of large collision systems has reached. The large amount of data collected in the past years allow to measure complex observables sensitive enough to provide unprecedented constraints to parameters of models aiming to describe the deconfined matter created in heavy-ion collisions. The most prominent examples presented here are the system and energy dependence of longitudinal flow vector decorrelations, differential studies of flow fluctuations, and measurements of non-linear response coefficients of harmonics up to very high orders and at variety of collision energies. Nevertheless, further studies, ideally using observables with exclusive sensitivity to either initial conditions, or the transport coefficients (especially the $\eta/s$ or $\zeta/s$), are still desired to reach a complete understanding of the deconfined medium created in heavy-ion collisions.

The origin of the collective effects observed in small collision systems is still elusive. Measurements driven by the fluctuating initial geometry suggest a large influence from final state effects, although it does not provide a clear answer on whether these are manifestation of a small fluid-like medium, or rather resemble a dilute system with just few parton scatterings. Nevertheless, contributions from initial state correlations should not be neglected in our considerations. In spite of the impressive advancements in both experimental and theoretical fields, we are still not able to conclude on what is the relative contribution of initial and final state effects to our experimental observations. This is an area of active development and hopefully some answers will be presented at the next edition of the Quark Matter conference.

\section{Acknowledgements}
\label{sec:acknowledgements}

I would like to express gratitude to the organisers of the Quark Matter 2019 conference for excellent scientific program, and for giving me the opportunity to summarise the most important recent developments in the field.

This work was supported by the project Centre of Advanced Applied Sciences with the number: \\CZ.02.1.01/0.0/0.0/16-019/0000778. Project Centre of Advanced Applied Sciences is co-financed by European Union.





\bibliographystyle{h-elsevier}
\bibliography{bibliography}







\end{document}